\documentclass[12pt,preprint]{aastex}

\usepackage{ulem}

\slugcomment{ApJ Letter accepted, December 1$^{\rm st}$ 2009}
\shorttitle{End of the WD CS in NGC 2158}
\shortauthors{Bedin et al.}

\begin{document}

\def\subr #1{_{{\rm #1}}}


\title{The bottom of the white dwarf cooling sequence 
        in the old open cluster NGC~2158\altaffilmark{1}}

\altaffiltext{1}{ Based on observations with the NASA/ESA {\it Hubble
Space Telescope}, obtained at the Space Telescope Science Institute,
which is operated by AURA, Inc., under NASA contract NAS 5-26555,
under GO-10500.}

\author{ L.\ R.\ Bedin\altaffilmark{2},  
         M.\ Salaris\altaffilmark{3}, 
         I.\ R.\ King\altaffilmark{4}.
         G.\ Piotto\altaffilmark{5},  
         J.\ Anderson\altaffilmark{2}, and 
         S.\ Cassisi\altaffilmark{6}. 
         }

\altaffiltext{2}{Space Telescope Science Institute, 3800 San Martin 
Drive, Baltimore, MD 21218; [bedin; jayander]@stsci.edu}

\altaffiltext{3}{Astrophysics Research Institute, Liverpool John Moores
University, 12 Quays House, Birkenhead, CH41 1LD, UK; ms@astro.livjm.ac.uk}

\altaffiltext{4}{Department of Astronomy, University of Washington,
Box 351580, Seattle, WA 98195-1580; king@astro.washington.edu}

\altaffiltext{5}{Dipartimento di Astronomia, Universit\`a di Padova,
Vicolo dell'Osservatorio 2, I-35122 Padova, Italy;
giampaolo.piotto@unipd.it}

\altaffiltext{6}{INAF-Osservatorio Astronomico di Collurania,
via M. Maggini, 64100 Teramo, Italy;
cassisi@oa-teramo.inaf.it}

\begin{abstract}
We use 10 orbits of Advanced Camera for Surveys observations to reach
the end of the white dwarf cooling sequence in the solar-metallicity
open cluster NGC 2158. Our photometry and completeness tests show that
the end falls at magnitude $m_{\rm F606W} = 27.5 \pm 0.15$, which
implies an age between $\sim$1.8 and $\sim$2.0~Gyr, consistent with
the age of $1.9 \pm 0.2$ Gyr obtained from fits to the main-sequence
turn-off.  The faintest white dwarfs show a clear turn toward bluer
colors, as predicted by theoretical isochrones.
\end{abstract}


%
\section{Introduction}
%
The recent discovery of an unexpectedly bright peak in the white dwarf
(WD) luminosity function (LF) of the metal-rich open cluster NGC 6791
(Bedin et al.\ 2005a, 2008a,b) has raised questions about the physical
processes that rule the formation of WDs and their cooling phases.
The aim of the present work is to investigate the WD cooling sequence (CS)
of another open cluster, NGC 2158, which has solar metallicity
(Jacobson, Friel, \& Pilachowski 2009) and is 4 times younger than NGC
6791 (Carraro et al.\ 2002), but somewhat less massive (as can be seen
from their images in the Sky Survey).  Our purpose is to extend our
knowledge of the dependence of WD LFs on cluster age and metallicity.

%
\section{Observations, Measurements, and Selections}
%
All data were collected with the wide field channel (WFC) of the
Advanced Camera for Surveys (ACS) at the focus of the {\it Hubble Space
Telescope} ({\it HST}) under program GO-10500 (PI:\ Bedin).  Data were
collected between October 24, 2005, and November 9, 2006, and consist of
1$\times$0.5s $+$ 4$\times$20s $+$ 10$\times$1145--1150s images in
filter F606W, and 1$\times$0.5s $+$ 4$\times$20s $+$
10$\times$1130--1150s in F814W.  Figure 1 shows a stacked image of the
region that we studied, after removal of cosmic rays and most of the
artifacts.  To study the WDs we used only the long exposures, in which
the average sky background is high enough that there are no CTE
problems.  The shortest exposures were used to study the brighest
and evolved cluster members. 

Photometry and relative positions were obtained with the software
tools described by Anderson et al.\ (2008).  In addition to solving
for positions and fluxes, we also computed two important diagnostic
parameters: The image-shape parameter RADXS is the fraction of light
that a source has outside the predicted PSF; it is very useful for
eliminating the faint blue galaxies that tend to plague studies of
WDs.  The sky-smoothness parameter rmsSKY is the rms deviation from
the mean, for the pixels in an annulus from 3.5 to 8 pixels from each
source.  As discussed in Bedin et al.\ (2008a), rmsSKY is invaluable
in measuring a more effective completeness than has been used in most
previous studies.

The photometry was calibrated into the WFC/ACS Vega-mag system
following the procedures given in Bedin et al.\ (2005b), and using
encircled energy and zero points given by Sirianni et al.\ (2005).  We
will use for these calibrated magnitudes the symbols $m_{\rm F606W}$
and $m_{\rm F814W}$.

Artificial-star (AS) tests were performed using the procedures described
by Anderson et al.\ (2008).  In the present program we chose them to
cover the magnitude range $24 < m_{\rm F606W} \leq 30$, with colors that
placed them on the WD sequence.  These played an important role in
showing us what selection criteria should be used for the real stars;
the quality of our results depends very much on making a good selection,
whose details are shown in Fig.\ 2.  (See Bedin et al.\ 2009 for a
detailed description of the selection procedures.)  Panels whose labels
are unprimed refer to real stars, while primed labels refer to the
artificial stars.

First, we required that a star appear in at least six long exposures in
each filter; panels e and e$^\prime$ show these selections, with the
cluster center marked by a $\star$.  Next, we used the AS to show what
combinations of magnitude and RADXS are acceptable for valid star images
(panel a$^\prime$), and we drew the red lines to isolate the acceptable
region.  We then drew
{\it these very same} 
lines in panel a, to separate the real stars from 
blends and probable galaxies.  
Note that the tail of objects on the right side in the AS panels is
produced by star-star blends that our simultaneous-fitting routine was
not able to separate into two components. These should certainly be
eliminated from our photometry lists (both real and AS).
We went through similar steps for the rmsSKY parameter
(panels b$^\prime$ and b).  (This step eliminated very few stars, but
was quite invaluable for the completeness estimate that we are about
to describe.)  Finally we plotted CMDs and drew dividing lines in a
similar way, to isolate the white dwarfs (panels c$^\prime$ and c).

Our final step was to deal with completeness, a concept that appears in
two different contexts: (1) The observed numbers of stars must be
corrected for a magnitude-dependent incompleteness.  (2) It is customary
to choose the limiting magnitude at the level where the sample is
50\% complete.  In a star cluster these two aims need to be treated in
quite different ways.  (1) To correct for incompleteness, we use the
traditional ratio of AS recovered to the number inserted.  (2) For the
limit to which measures of faint stars are reliable, however, the
50\%-completeness level needs to be chosen in a quite different way,
because in a crowded star cluster more faint stars are lost in the
brightness fluctuations around the bright stars than are lost to the
fluctuations of the sky background.  The completeness measure that we
should use to set the 50\% limit is therefore the recovered fraction of
AS {\it only} among those AS whose value of rmsSKY, as a function of
magnitude, indicates that it {\it was possible to recover the star at
all\/}.  [For a more detailed discussion, see Sect.\ 4 of Bedin et al.\
(2008a)].

The distinction between the two completeness measures is illustrated
in panel d of Fig.\ 2, where the black crosses and line show the
overall completeness, while the blue squares and line show the
completeness statistic that takes rmsSKY into account; its 50\%
completeness level is at $m_{\rm F606W}=28.95$, rather than the 28.26
with which the traditional statistic would have left us.  To emphasize
the contrast, the red circles and line show the fraction of the area
in which faint stars could have been found at all.

%
\section{Comparison with Theory}
%

The comparison with theory has three parts: 
(a) determination of a main-sequence (MS) turn-off (TO) age, 
(b) comparison between the observed WD sequence and theoretical isochrones, and 
(c) determination of the age that is implied by the WD LF.

For the fitting of the MSTO we used the BaSTI scaled-solar models and
isochrones, including convective-core overshooting (Pietrinferni et
al.\ 2004), but we needed first to choose a value of the metallicity.
The literature of NGC 2158 had offered a wide range of [Fe/H] values
($\sim-$0.2 to $\sim-$0.9) and of $E(B-V)$ ($\sim$0.4 to $\sim$0.6),
as recently summarized by Jacobson et al.\ (2009).  Those authors
presented the first high-resolution spectroscopic analysis of NGC
2158, and arrived at ${\rm [Fe/H]}=-0.03\pm0.14$.  From a cubic
interpolation among the isochrones in the BaSTI database, we derived
scaled-solar isochrones for that [Fe/H] value, and determined the
cluster reddening and distance modulus, as follows.  We first allowed
for the effect of extinction, as described by Bedin et al.\ (2009) for
the globular cluster M4, and we then estimated the distance modulus,
reddening, and age of the cluster by matching isochrones
simultaneously to the mean color and magnitude of the red clump (to
fix distance and extinction --- see the box in Fig.\ 3) and to the
location of the TO region (to fix the age).

The upper-right panel of Fig.\ 3 shows isochrones that bracket the
observed TO luminosity, two with overshooting and two without.  It is
clear that no good fit to the MSTO region can be obtained without
overshooting.  The isochrones that fit the red clump and the MSTO also
give a good fit to the MS.  For ${\rm [Fe/H]}=-0.03$ we obtain
$A_V=1.3$ [corresponding to $E(B-V)=0.42$], $(m-M)_0=12.98$, and an
age between 1.75 and 2.0 Gyr.  This age is consistent with previous
results by Carraro et al.\ (2002) and Salaris et al.\ (2004) with
different methods, models, and [Fe/H], while the values of reddening
and distance modulus agree with results by Grocholski \& Sarajedini
(2002) based on a study of the red clump in the $K$ vs.\ $(J-K)$ CMD.

To estimate the error in the fit of the sparsely populated red clump
(only 12 stars), we synthesized from each isochrone a rich CMD of this
region (several thousand stars), and compared these theoretical
representations of the red clump with the observed one. The 1-$\sigma$
spreads in the fit are $\pm$0.05 mag in the mean $m_{\rm F606W}$, and
$\pm$0.02 mag in the mean color.

A total uncertainty of $\pm$0.06 mag for $(m-M)_0$ has been estimated
by adding in quadrature the $\pm$0.05 mag uncertainty in the mean
$m_{\rm F606W}$ of the observed red clump, and a $\pm$0.03 mag
contribution due to the change of the mean $m_{\rm F606W}$ of the
isochrone red clump when [Fe/H] is changed by its $\pm$0.14 dex
uncertainty.  (We did this for a fixed age of 2 Gyr; the mean
magnitude of the isochrone red clump changes negligibly for age
changes of a few hundred Myr around that value.)

To estimate the error in $A_V$, we first added in quadrature the
$\pm$0.02 mag uncertainty in the mean color of the observed red clump,
and a $\pm$0.03 mag uncertainty in the mean color of the isochrone red
clump, the latter corresponding to the uncertainty in [Fe/H].  The
combined color uncertainty was then converted to a $\pm$0.09 mag
uncertainty in $A_V$, by using the Cardelli et al.\ (1989) extinction
law, with $R_V=3.1$.

As for the age, we estimated its total uncertainty by adding four
separate contributions in quadrature:\ first a $\pm$0.125 Gyr
uncertainty in the fit itself, even at fixed values of [Fe/H],
distance, and extinction, and then contributions from allowing each
one of those three to vary while holding the other two constant ---
$\pm$0.08 Gyr from [Fe/H], $\pm$0.06 Gyr from the distance modulus,
and $\pm$0.08 Gyr from the extinction.  Our final value of the MSTO
age is $1.9\pm0.2$ Gyr.

We move now to the white dwarfs.  For them we found it preferable to
work with different data samples for the shape of the WD CS and for
the WD LF.  The best WD CMD came from a radial selection (blue circle
in Fig.\ 2) that excludes the crowded cluster center; this leads to a
cleaner WD sequence for the comparison with isochrones (top-left panel
of Fig.\ 3).  Such a selection, however, would drastically reduce in
the numbers of stars in our WD LF, where narrowing of the sequence is
of no benefit; we therefore used the entire sample (shown in panel c
of Fig.\ 2) for the fitting of the LF.

One more point to note is that in the LF we concentrate only on the
peak at the faint end, because dynamical evolution in the cluster has
distorted the detailed shape of the LF in a way that is too
complicated to model.
This is suggested by the binary sequence that can be seen to the right
of the MS, also in panel c of Fig.\ 2.  Unlike the smooth distribution
of binaries that is seen in many globular clusters, the binary
sequence here is sharply peaked at the red edge that corresponds to
equal-mass binaries.  Binaries of unequal mass ratio have a lower mass
than those of equal mass, and have been largely removed by the
mass-dependent escape of stars.  The effect on the WD LF is very hard
to estimate in detail, but there is no effect on the location of the
peak at the faint end, which is what we use for estimating the cluster
age.

For WD theory we used the carbon-oxygen WD models by Salaris et al.\
(2000), transformed into the ACS Vega-mag system as described in Bedin
et al.\ (2005a,b).  The top left panel of Fig.\ 3 compares WD
isochrones with the observed WD sequence.  We used isochrones from our
H-atmosphere WD models, along with the initial-final-mass relationship
by Salaris et al.\ (2009), and the progenitor lifetimes for ${\rm
  [Fe/H]}=-0.03$, obtained again by interpolation among the values
from the BaSTI database.  Isochrones for 1.75 and 2.0 Gyr are
overplotted on the observed WD sequence, using our derived values
$A_V=1.3$ and $(m-M)_0=12.98$.
In the more vertical part of the WD sequence, where the errors are
small, they fit well, and they turn to the blue at the right magnitude. 

The WD cooling track for a 0.61 $M_{\odot}$ WD with He envelope is
also shown.  Its mass is close to the mass ($\sim$0.65 $M_{\odot}$) of
objects evolving along the bright linear part of the H-atmosphere WD
isochrone, and we take it as representative of the non-DA population
that may possibly contaminate the observed cooling sequence.

We also show the cooling track of the 0.465$M_{\odot}$ He-core WDs
that we used in our previous work on NGC 6791 (Bedin et al.\ 2008a) to
estimate the location of He-core WDs --- if they exist in the cluster
--- compared with the CO-core sequence. The maximum
electron-degenerate He-core mass for NGC 2158 stars is
$\sim0.45\,M_{\odot}$ for the objects currently at the tip of the red
giant branch (total masses of the order of 1.7--1.8 $M_{\odot}$);
therefore the He-core WDs --- in the absence of episodes of mass
accretion from He-core WD companions --- have to populate a CMD region
to the red of the 0.465 $M_{\odot}$ model that is shown.  We conclude
from Fig.\ 3 that the contamination of the observed cooling sequence
by He-core WDs has to be negligible.

The lower panel of Fig.\ 3 compares the observed WD LF in F606W with theory.  
The observed sample contains 138 WD candidates, which become  
183 after correction for incompleteness. 
We show the two ages, 1.75 and 2.0 Gyr, that bracket the MSTO age.
Our theoretical LFs come from H-atmosphere WD isochrones that were
produced by using a Monte Carlo method to synthesize a WD sequence, as
described in Bedin et al.\ (2008b).  The simulations minimize the
effect of random fluctuations of star counts, by using $\sim$1000
times as many WDs as the observed sequence, and include appropriate
photometric errors as found from the data-reduction procedure.  We
used a Salpeter mass function for the WD progenitors, and included the
effect of incompleteness by randomly removing stars according to the
completeness curves described in Section 2.

Given the presence of a clear binary sequence that parallels the the
MS, we included in our synthetic WD sequences a 20\% fraction of
unresolved WD+WD binary systems.  (See Bedin et al.\ 2008b for
details.).  Neither the inclusion of WD+WD binaries nor the precise
choice of the progenitor mass function affects the position of the LF
cutoff for a given age.  For the plots of Fig.\ 3 the total number of
stars in the synthetic WD sequences has been rescaled to match the
observed number of objects in the magnitude interval $24.5 \le m_{\rm
F606W} \le 26.0$.

The observed and theoretical LFs are clearly consistent.  In
particular, if we locate the cut-off of the observed LF at the
magnitude of the faintest bin whose star counts are above zero by
3$\sigma$ ($m_{\rm F606W}=27.500\pm0.125$), the two LFs for 1.75 and
2.0 Gyr represent lower and upper limits for the age inferred from the
WDs, in nice agreement with the result from the MSTO.  We note also
that the uncertainties in the distance modulus and in the metallicity
of the progenitors have a negligible effect ($\sim\pm 50$ Myr) on this
age range.

The overall shapes of the observed LFs in both filters are very
similar to their theoretical counterparts, but we refrain at this
stage from attempting a more quantitative fit (or even a
bi-dimensional fit of the WD distribution in the CMD) for several
reasons: The exact shape of the WD LF (and of the WD sequence in the
CMD) depends on a number of parameters in addition to the cluster age,
e.g., the initial-final-mass relationship, the mass function of the
progenitors, the relative fraction of H-and He-atmosphere objects, the
fraction of WD+WD binaries (and their mass distribution) and, very
importantly (as explained earlier in this section), dynamical
evolution of the cluster, which selectively depletes the WDs according
to their mass and their time of formation.  The observed position of
the LF cut-off, however, which determines our WD age, should have been
affected very little, if at all, by the dynamics of the cluster, for
the following reason.  The WD isochrones predict that for ages around
the MSTO age, the brighter WDs should have nearly a single mass,
whereas the 0.25-mag bin at the LF cutoff --- i.e., at the blue-turn
at the bottom of the CS --- should be populated by objects spanning
almost the whole predicted WD mass spectrum.  A mass-dependent loss
mechanism would therefore alter the number distribution of WDs around
the blue-turn, but it should have no effect on the observed location
of the cut-off in the WD LF.

We close this section by addressing briefly the issue of WDs that have
an atmosphere of helium rather than hydrogen.  For the temperature
range of the NGC 2158 cooling sequence, the typical number ratio of
He- to H-atmosphere WDs in the field is of the order of 1:4 (e.g.,
Tremblay \& Bergeron 2008), but it is not clear whether there is a
dearth of He-atmosphere WDs in star clusters compared to the field
(see, e.g., Strickler et al.\ 2009, Davis et al.\ 2009).  In spite of
this uncertainty, at the luminosities of the NGC 2158 WDs the
He-atmosphere models have cooling ages very similar to those of
H-atmosphere models at the same mass (see, e.g., Hansen 1999, Salaris
2009).  If the observed cooling sequence includes He-atmosphere WDs,
even in substantial number, their effect on the ages estimated from
the position of the LF cutoff would be negligible.

\acknowledgements
We warmly thank an anonymous referee for useful comments.
J.A.\ and I.R.K.\ acknowledge support from STScI grant GO-10500. 
GP acknowledges support by PRIN2007 and ASI under the program ASI-INAF
I/016/07/0. 
%
%

%
%


\begin{figure}
\epsscale{1.00}
\plotone{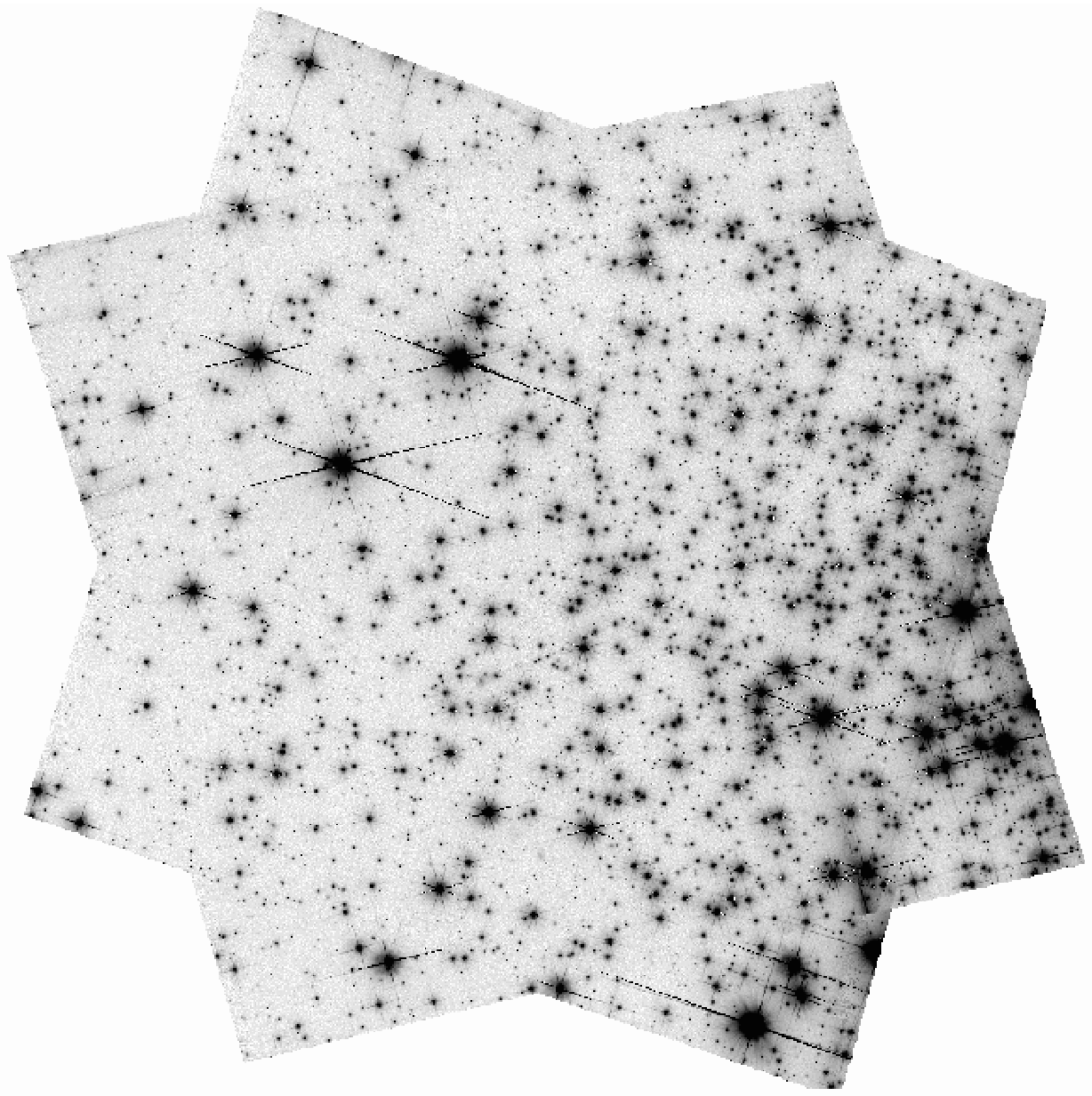}
\caption{Stacked image of deep exposures in filter F814W.}
\end{figure}

\begin{figure}
\epsscale{1.00}
\plotone{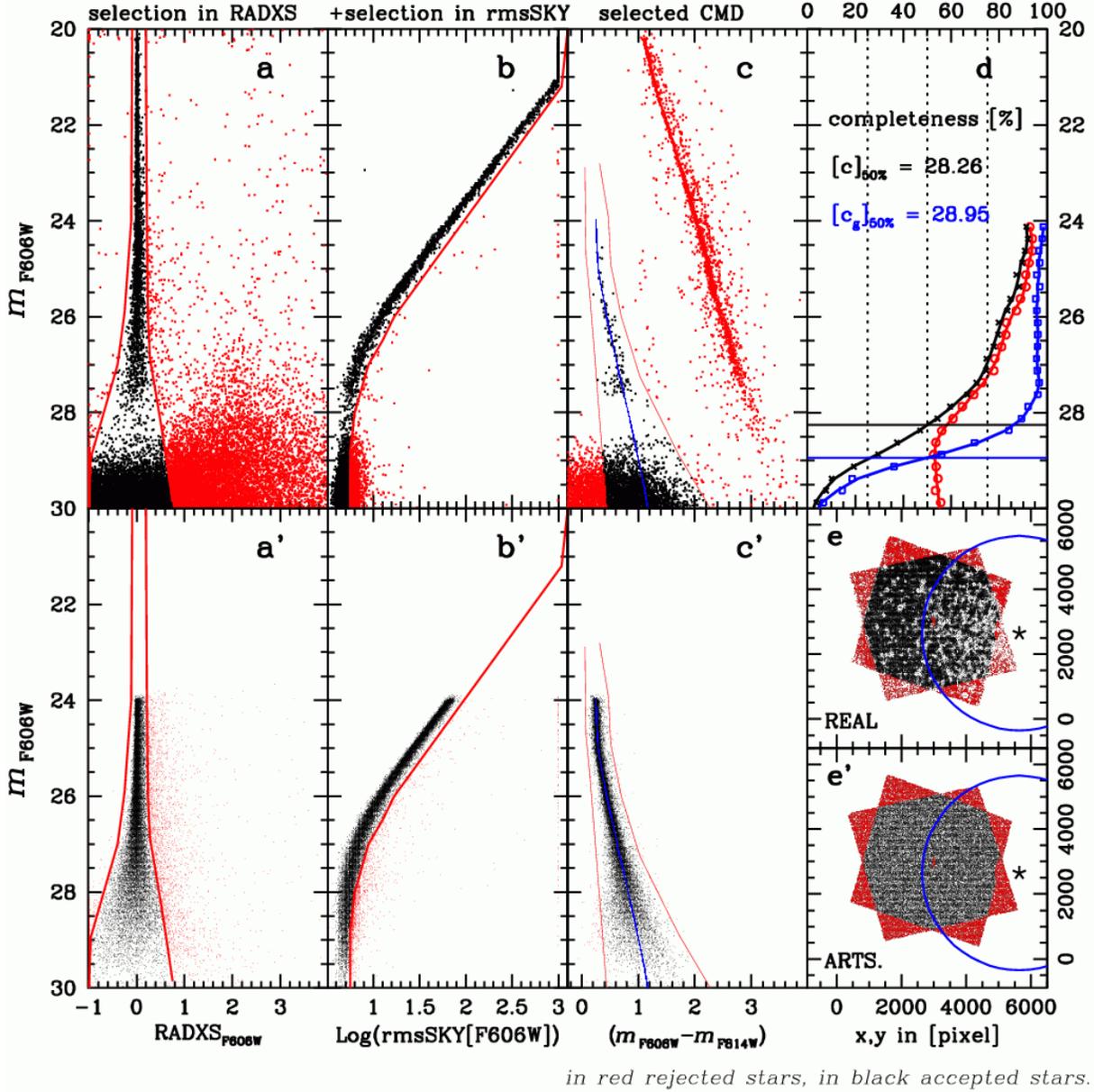}
\caption{
(a) and (a$^\prime$):\ The parameter RADXS as a function of magnitude,
for real stars and AS.  Objects that we consider to be real stars lie
between the two red lines.  (b) and (b$^\prime$):\ The parameter
rmsSKY as a function of magnitude.  Stars that lie on an acceptable
background lie to the left of the red line.  (c) and (c$^\prime$):\
The CMD, with red lines delineating the WD region.  
The blue lines show the fiducial line along which  
the AS were added. 
(d):\ Black points and line show the traditional ``completeness'' that
ignores the rmsSKY criterion, while the blue points and line show the
completeness in the regions in which faint stars can be detected and
measured.  The red circles and line show the fraction of the total that
this measurable area comprises.  Panels (e) and (e$^\prime$):\ Spatial
distribution of the real and AS detections. In black, objects that fell
in at least 6 deep images in both F606W and F814W.  The assumed cluster
center is marked with a $\star$.  The blue circles identify the radius
outside of which the more restricted WD sample was taken (see Sect.\
3).}
\end{figure}

\begin{figure}
\epsscale{1.00}
\plotone{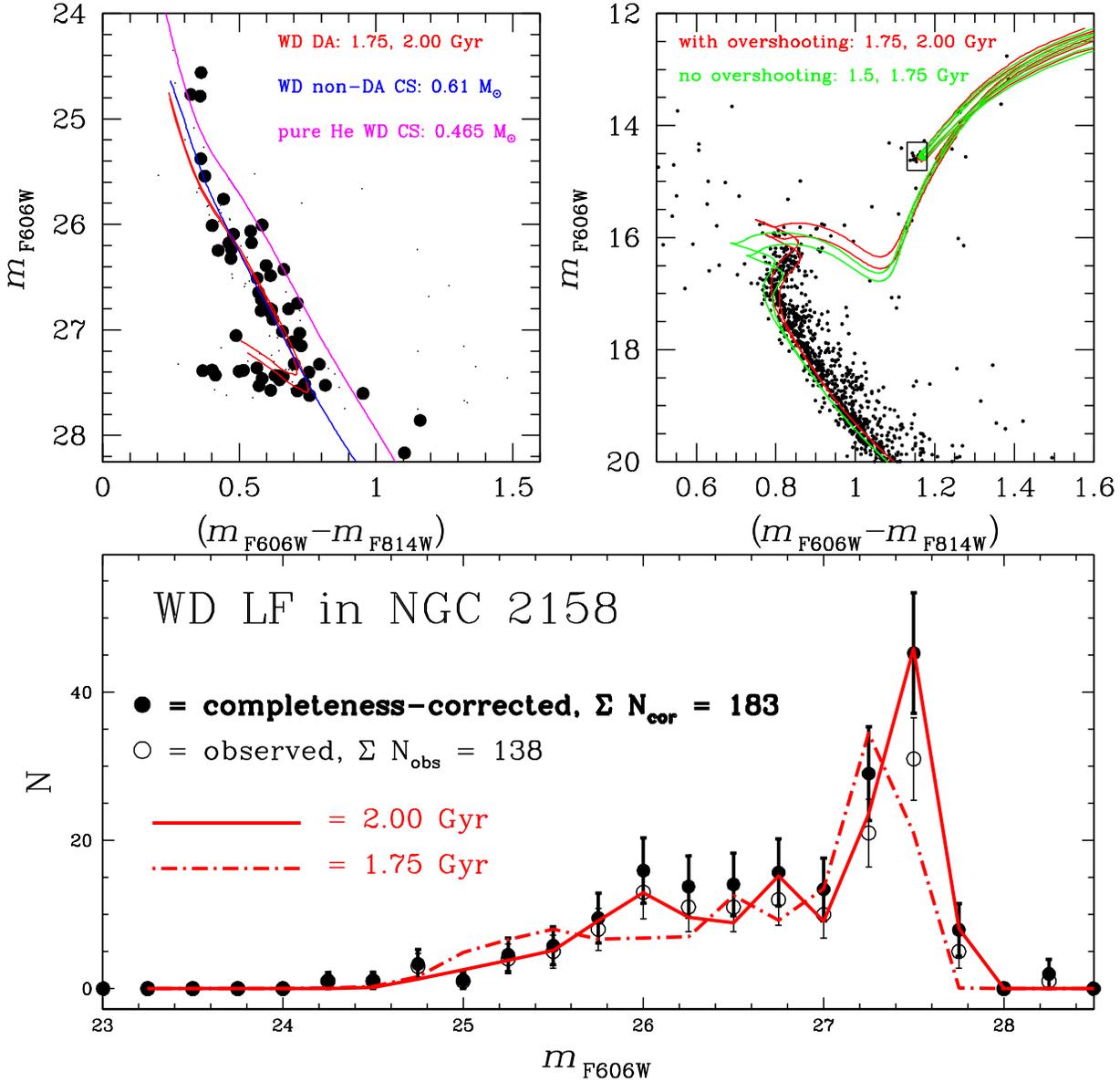}
\caption{
({\it upper left:}) The observed cooling sequence compared to
theoretical H-atmosphere WD isochrones (red lines) for 1.75 and 2.0 Gyr,
shifted in color and magnitude according to the distance modulus and
extinction estimated from fits to the MS, its TO, and the red clump.
The blue line is the cooling track of a 0.61$M_{\odot}$ He-atmosphere
WD, while the magenta line represents a 0.465$M_{\odot}$ He-core WD.
({\it upper right:}) Fit of theoretical isochrones to the cluster CMD,
from the MS to the red clump, for ages of 1.75 and 2.0 Gyr.  ({\it
bottom:}) comparison between the observed WD LF in F606W and its
theoretical counterpart, for 1.75 and 2.0~Gyr.  (See text for details).
}
\end{figure}

\begin{figure}
\epsscale{1.00}
\plotone{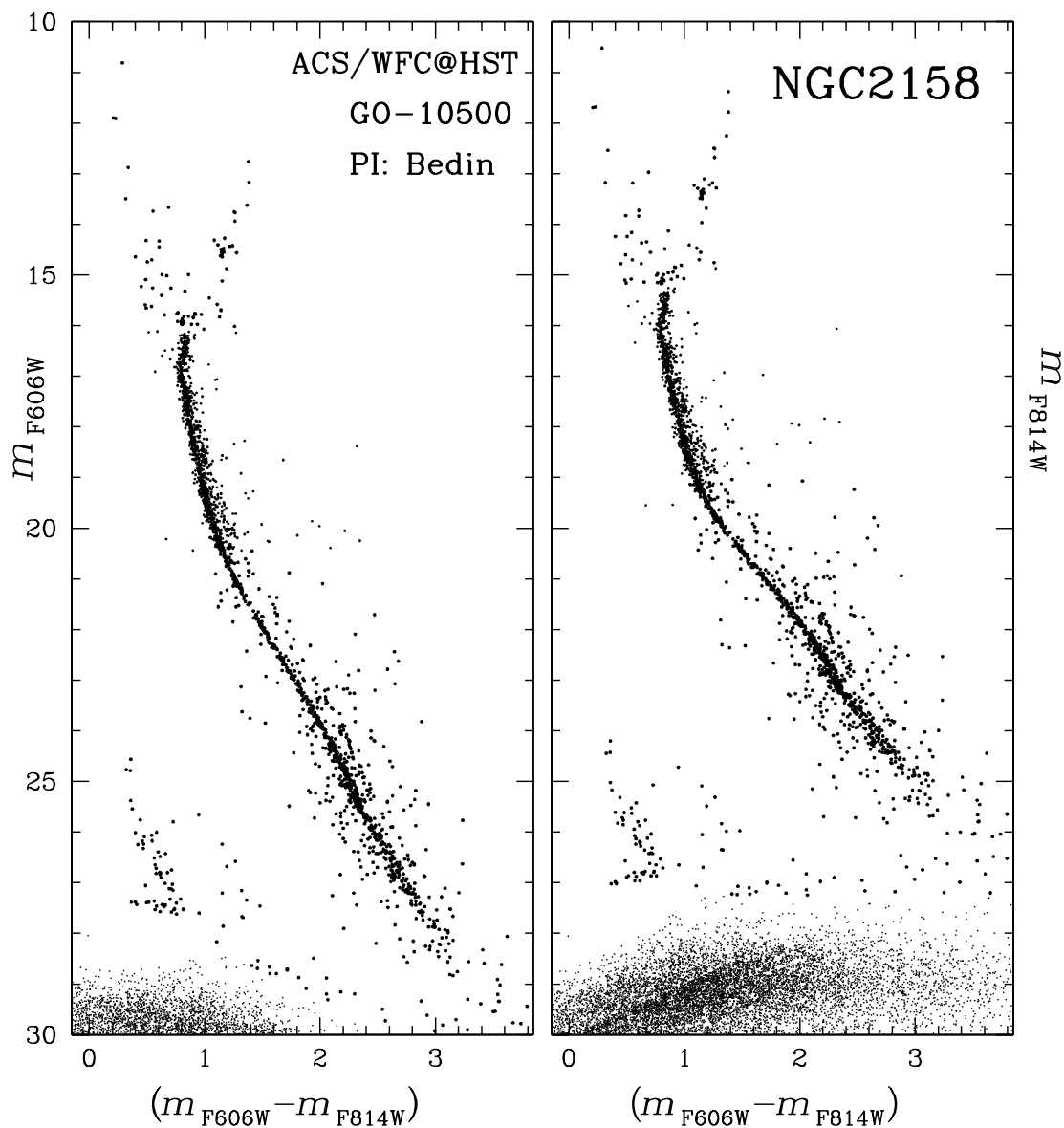}
\caption{
{\bf (Bonus-material only for the astro-ph version of the draft):} 
The two panels show the entire CMDs, obtained including short-,
middle- and long-exposures.  Note that photometries derived from
images of different exposure times are subject to different
selections.  Objects measured in the short (0.5 s) and long (1200 s)
exposures are marked with larger filled circles than those measured in
middle exposures. Small dots are considered non-significant 
detections and indicate the floor-noise level in the background.
}
\end{figure}

\end{document}